\documentstyle[sprocl]{article}
\title{ON A CANONICAL FORMULATION OF STRING THEORY IN MASSIVE
BACKGROUND FIELDS}
\author{I.L. BUCHBINDER, V.D. PERSHIN, G.B. TODER}
\address{Department of Theoretical Physics,
Tomsk State Pedagogical University,
Tomsk 634041, Russia}
\begin{document}
\maketitle\abstracts{
We propose a method of constructing a gauge invariant canonical
formulation for non-gauge classical theory which depends on a
set of parameters. Requirement of closure for algebra of
operators generating quantum gauge transformations leads to
restrictions on parameters of the theory. This approach is then
applied to bosonic string theory coupled to massive background
fields. It is shown that within the proposed canonical
formulation the correct linear equations of motion for
background fields arise.}

The BFV method~\cite{BFV} is the most general realization of
canonical quantization procedure providing a natural and
consistent approach to construction of quantum models in
theoretical physics. In this paper we discuss one of its yet
unexplored aspects arising from bosonic string theory coupled to
background fields~\cite{callan}.

A crucial point of string models is requirement
of conformal invariance at the quantum level. It leads to
restrictions on spacetime dimension in the case of free string
theory and to effective equations of motion
for massless background fields in the case of string theories
coupled to background~\cite{callan}.
According to the prescription  generally accepted in
functional approaches to string theory~\cite{Pol} dynamical
variables should be treated in different ways. Namely,
functional integration is carried out only over string
coordinates while components of
two-dimensional metric are considered as
external fields. This approach can also be applied to string theory
interacting with massive background fields which is not
classically conformal invariant~\cite{ovrut}.
As was shown in\cite{buchper} it gives rise to effective
equations of motion for massive background fields.
Unfortunately, in the case of closed string theory
covariant approaches did not reproduce the full set of correct
linear equations of motion for massive background
fields~\cite{buchper,ellw} and so there exists a problem of deriving correct
massive background fields equations.

Moreover, from general point of view the requirement of quantum
Weyl invariance of string theory with massive background fields
means that a non-gauge classical theory depending on a set of
parameters is used for constructing of a quantum theory that is
gauge invariant under some special values of the parameters.
As we consider canonical approach to be the only completely
consistent method for constructing quantum theories so the
general problem arising from string theories is how to describe
in terms of canonical quantization construction of gauge
invariant quantum theory starting with a classical theory
without this invariance.

Due to the general BFV method one
should construct hamiltonian formulation of classical theory
and define fermionic functional $\Omega$
generating algebra of gauge transformations and bosonic
functional $H$ containing information of theory dynamics.
Quantum theory is consistent provided that the operator $\hat\Omega$
is nilpotent and conserved in time. The corresponding analysis
for bosonic string coupled to massless background fields was
carried out in the paper~\cite{buch91}.
In the case of string theory interacting with massive background
fields classical gauge
symmetries are absent and it is impossible to construct
classical gauge functional $\Omega$. In this paper we propose a
prescription allowing for some models to construct quantum
operator $\hat\Omega$ starting with a classical theory without
first class constraints. Then we apply it to the theory of closed
bosonic string coupled to masssive background fields.

Consider a system described by a hamiltonian
\begin{equation}
H=H_{0}(a)+\lambda^{\alpha}T_{\alpha}(a)
\label{Hgen}
\end{equation}
where $H_0(a)=H_0(q,p,a)$, $T_{\alpha}(a)=T_{\alpha}(q,p,a)$ and
$q$, $p$ are canonically conjugated dynamical variables;
$a=a_i$ and $\lambda^{\alpha}$ are external parameters of the
theory. We suppose that $T_{\alpha}$ are some functions of the form
$T_{\alpha}=T^{(0)}_{\alpha}+T^{(1)}_{\alpha}$
and closed algebra in terms of Poisson brackets is formed by
$T^{(0)}_{\alpha}$, not by $T_{\alpha}$:
\begin{equation}
\{ T^{(0)}_{\alpha}(a),T^{(0)}_{\beta}(a) \}  =
T^{(0)}_{\gamma}(a)U^{\gamma} _{\alpha\beta}(a), \quad
\{ H_0(a),T^{(0)}_{\alpha}(a) \}  =
T^{(0)}_{\gamma}(a)V^{\gamma} _{\alpha}(a)
\end{equation}
Such a situation may occur, for example, if
$T^{(0)}_{\alpha}$ correspond to a free gauge invariant
theory and $T^{(1)}_{\alpha}$ describe a perturbation
spoiling gauge invariance. At the quantum level both the algebras of
$T^{(0)}_{\alpha}(a)$ and $T_{\alpha}(a)$ are not closed in
general case.

We define quantum operators $\Omega$ and  $H$ as follows:
\begin{equation}
\Omega =
c^{\alpha}T_{\alpha}(a)-{1\over2}U^{\gamma}_{\alpha\beta}(a)
:{\cal P_{\gamma}} c^{\alpha} c^{\beta}:, \quad
H  = H_0 (a) + V^{\gamma}_{\alpha}(a) :{\cal P_{\gamma}}
c^{\alpha}:
\label{Omega}
\end{equation}
where $:\quad:$ stands for some ordering of ghost fields.
In general case $\Omega^2 \neq 0$ and $d\Omega/dt \neq 0$.
However, if there exist some specific values of parameters $a$ that make
the operator $\Omega$ to be nilpotent and conserved  then the corresponding
quantum theory is gauge invariant. Thus it is possible
to construct a quantum theory with given gauge
invariance that is absent at the classical level.

As  an example where the described procedure really
works we consider closed bosonic string theory
coupled with background fields of tachyon and of the first
massive level in linear approximation.\footnote{An adequate
treatment of non-linear (interaction)
terms is known to demand non-perturbative methods~\cite{Das}.}
The theory is described by the classical action
$$
S={}-{1\over 2\pi\alpha'} \int d^{2}\sigma\,
  \sqrt{-g} \bigl\{{1\over 2} g^{ab}\partial_{a} x^{\mu} \partial_{b}
  x^{\nu} \eta_{\mu\nu} + Q(x)
$$
$$
 {}+  g^{ab} g^{cd} \partial_a x^\mu \partial_b x^\nu
          \partial_c x^\lambda \partial_d x^\kappa
        F^1_{\mu\nu\lambda\kappa}(x)
  + g^{ab}\varepsilon^{cd} \partial_a x^\mu \partial_b x^\nu
          \partial_c x^\lambda \partial_d x^\kappa
      F^2_{\mu\nu\lambda\kappa}(x)
$$
\begin{equation}
{}+ \alpha' R \, g^{ab} \partial_a x^\mu \partial_b x^\nu
      W^1_{\mu\nu}(x) +
  \alpha' R \, \varepsilon^{ab} \partial_a x^\mu \partial_b x^\nu
      W^2_{\mu\nu}(x)
  +  \alpha'^2 R \, R \, C(x) \bigr\},
\label{S}
\end{equation}
$\sigma^a=(\tau,\sigma)$ are coordinates on string world sheet,
$R$ is scalar curvature of $g^{ab}$,
$\eta_{\mu\nu}$ is Minkowski metric of
$D-$dimensional spacetime,  $Q$ is tachyonic field and $F$,
$W$, $C$ are background fields of the first massive level.
As was shown in~\cite{buchper} all other
possible terms with four two-dimensional derivatives in
classical action are not essential and string interacts with
background fields of the first massive level only by means of
the terms presented in (\ref{S}).

Components of two-dimensional metric $g_{ab}$ should be
considered as external fields, otherwise the classical equations
of motion $\delta S/\delta g_{ab}=0$ would be fulfilled only for
vanishing background fields. Such a treatment
corresponds to covariant methods where functional integral
is calculated only over $x^\mu$ variables.
After the standard parametrization of the metric
\begin{eqnarray}
g_{ab}& = &e^{\gamma}\left(\begin{array}{cc}
\lambda^{2}_{1}-\lambda^{2}_{0} & \lambda_{1} \\
\lambda_{1}  & 1 \end{array}\right)
\end{eqnarray}
the hamiltonian in linear approximation in background fields
takes the form $H=\int d\sigma \,(\lambda_{0}T_{0}+\lambda_{1}T_{1})$
where $T_0 = T_0^{(0)} + T_0^{(1)}$, $T_1 = T_1^{(0)}$ and
$T^{(0)}$ represent constraints of free string
theory forming closed algebra in terms of Poisson brackets.
In free string theory conditions
$T^{(0)}=0$ result from conservation of
canonical momenta conjugated to $\lambda$.
According to our prescription in string theory with massive
background fields  $\lambda$ can not be
considered as dynamical variables, there are no corresponding
momenta and conditions of their conservation do not appear.

The role of parameters $a$ is
played by background fields and conformal factor $\gamma$ and the
theory is of the type (\ref{Hgen}) with $H_0=0$.
Direct calculations up to terms linear in bakground fields show
that the operator $\Omega$ defined according to (\ref{Omega})
is nilpotent and conserved in this theory under the following conditions:
$$
D=26, \quad \gamma=const, \quad
(\partial^2 + 4/\alpha')Q=0, \quad
(\partial^2 - 4/\alpha')F_{\mu\nu,\lambda\kappa}=0 ,
$$
\begin{equation}
\partial^\mu F_{\mu\nu,\lambda\kappa}=0, \quad
\partial^\lambda F_{\mu\nu,\lambda\kappa}=0, \quad
F^\mu{}_{\mu,\lambda\kappa}=0, \quad
F_{\mu\nu,}{}^\lambda{}_\lambda=0.
\label{shell}
\end{equation}
The condition $\gamma=const$ means that
string world sheet should be flat $R=0$ and so
the background fields $W$ and $C$ disappear from the classical
action (\ref{S}). The eqs.(\ref{shell}) show
that first massive level is described by a tensor
of fourth rank which is symmetric and traceless in two pairs of
indices and transversal in all indices. This exactly corresponds
to the closed string spectrum and so our approach gives the full
set of correct linear equations for massive background fields.

The described example demonstrates a possibility to construct
canonical formulation of quantum theory invariant under gauge
transformations that are absent at the classical level. The
proposed method opens up a possibility for deriving interacting
effective equations of motion for massive and massless
background fields within the framework of canonical formulation
of string models and provides a justification of covariant
functional approach to string theory.

The authors are grateful to E.S.~Fradkin, P.M.~Lavrov,
R.~Marnelius, B.~Ovrut and I.V.~Tyutin for
discussions of some aspects of the paper. The work was supported
by International Scientific Foundation, grant No RI~1300, and
Russian Foundation for Fundamental Research, project No
96-02-16017.

\end{document}